\documentclass[aps,prl,twocolumn,superscriptaddress,showpacs]{revtex4-1}
\usepackage{graphicx}
\usepackage[T1]{fontenc}
\usepackage{multirow}
\usepackage{tabularx}
\usepackage{comment}
\usepackage{xcolor}
\usepackage{threeparttable}
\usepackage{bm}
\usepackage{lipsum}
\usepackage{setspace}
\usepackage{amsmath}
\usepackage{physics}

\begin{document}

\title{Spectroscopy of $^{52}$K}

\newcommand{\aatomki}{     \affiliation{HUN-REN Institute for Nuclear Research, HUN-REN ATOMKI, P.O. Box 51, Debrecen H-4001, Hungary}}
\newcommand{\abeijing}{    \affiliation{State Key Laboratory of Nuclear Physics and Technology, Peking University, Beijing 100871, China}}
\newcommand{\abnu}{\affiliation {Key Laboratory of Beam Technology and Material Modification of Ministry of Education, College of Nuclear Science and Technology, Beijing Normal University, Beijing 100875, China}}
\newcommand{\acaen}{       \affiliation{LPC Caen, Normandie Univ, ENSICAEN, UNICAEN, CNRS/IN2P3, F-14000 Caen, France}}
\newcommand{\acea}{        \affiliation{IRFU, CEA, Universit\'e Paris-Saclay, F-91191 Gif-sur-Yvette, France}}
\newcommand{\acns}{        \affiliation{Center for Nuclear Study, University of Tokyo, RIKEN campus, Wako, Saitama 351-0198, Japan}}
\newcommand{\aewha}{       \affiliation{Ewha Womans University, Seoul 03760, Korea}}
\newcommand{\aibs}{        \affiliation{Institute for Basic Science, Daejeon 34126, Korea}}
\newcommand{\agsi}{        \affiliation{GSI Helmholtzzentrum f\"ur Schwerionenforschung GmbH, Planckstr. 1, 64291 Darmstadt, Germany}}
\newcommand{\ahku}{        \affiliation{Department of Physics, The University of Hong Kong, Pokfulam, Hong Kong}}
\newcommand{\ainst}{       \affiliation{Institute for Nuclear Science \& Technology, VINATOM, 179 Hoang Quoc Viet, Cau Giay, Hanoi, Vietnam}}
\newcommand{\aipno}{       \affiliation{IPN Orsay, CNRS, Univ. Paris Sud, Univ. Paris-Saclay, F-91406 Orsay Cedex, France}}
\newcommand{\aijclab}{     \affiliation{Universit\'e Paris-Saclay, CNRS/IN2P3, IJCLab, F-91405 Orsay cedex, France}}
\newcommand{\akoeln}{      \affiliation{Institut f\"ur Kernphysik, Universit\"at zu K\"oln, D-50937 Cologne, Germany}}
\newcommand{\akth}{        \affiliation{Department of Physics, Royal Institute of Technology, SE-10691 Stockholm, Sweden}}
\newcommand{\alanzhou}{    \affiliation{Institute of Modern Physics, Chinese Academy of Sciences, Lanzhou 730000, China}}
\newcommand{\amadrid}{     \affiliation{Instituto de Estructura de la Materia, CSIC, E-28006 Madrid, Spain}}
\newcommand{\amsu}{        \affiliation{Department of Physics and Astronomy, Michigan State University, East Lansing, MI 48824-1321, United States}}
\newcommand{\aorsay}{      \affiliation{CSNSM, CNRS/IN2P3, Universit\'e Paris-Sud, F-91405 Orsay Campus, France}}
\newcommand{\aoslo}{       \affiliation{Department of Physics, University of Oslo, N-0316 Oslo, Norway}}
\newcommand{\ariken}{      \affiliation{RIKEN Nishina Center, 2-1 Hirosawa, Wako, Saitama 351-0198, Japan}}
\newcommand{\arikkyo}{     \affiliation{Department of Physics, Rikkyo University, 3-34-1 Nishi-Ikebukuro, Toshima, Tokyo 172-8501, Japan}}
\newcommand{\atitech}{     \affiliation{Department of Physics, Institute of Science Tokyo, 2-12-1 O-Okayama, Meguro, Tokyo, 152-8551, Japan}}
\newcommand{\atohoku}{     \affiliation{Department of Physics, Tohoku University, Sendai 980-8578, Japan}}
\newcommand{\atudarmstadt}{\affiliation{Institut f\"ur Kernphysik, Technische Universit\"at Darmstadt, 64289 Darmstadt, Germany}}
\newcommand{\aunal}{       \affiliation{Universidad Nacional de Colombia, Sede Bogot\'a, Facultad de Ciencias, Departamento de Física, Bogot\'a 111321, Colombia}}
\newcommand{\ajaver}{      \affiliation{Pontificia Universidad Javeriana, Facultad de Ciencias, Departamento de F\'isica, Bogot\'a, Colombia}}
\newcommand{\aut}{         \affiliation{Department of Physics, University of Tokyo, 7-3-1 Hongo, Bunkyo, Tokyo 113-0033, Japan}}
\newcommand{\azagreb}{     \affiliation{Ru{\dj}er Bo\v{s}kovi\'c Institute, Bijeni\v{c}ka cesta 54, 10000 Zagreb, Croatia}}
\newcommand{\abro}{        \affiliation{Laboratoire Kastler Brossel, Sorbonne Universit\'e, CNRS, ENS, PSL Research University, Coll\`ege de France, Case 74, 4 Place Jussieu, 75005 Paris, France}}
\newcommand{\APoves}{        \affiliation{Departamento de Fisica Teorica and IFT UAM-CSIC, Universidad Autonoma de Madrid, Spain}}
\newcommand{\FNowacki}{        \affiliation{Université de Strasbourg, CNRS, IPHC UMR 7178, F-67000 Strasbourg, France}}
\newcommand{\KOgata}{        \affiliation{Research Center for Nuclear Physics (RCNP), Osaka University, Ibaraki 567-0047, Japan}}
\newcommand{\KOgataS}{        \affiliation{Department of Physics, Kyushu University, Fukuoka 819-0395, Japan}}
\newcommand{\KYoshida}{        \affiliation{Advanced Science Research Center, Japan Atomic Energy Agency, Tokai, Ibaraki 319-1195, Japan}}
\newcommand{\afair}{      \affiliation{Helmholtz Forschungsakademie Hessen für FAIR (HFHF), GSI Helmholtzzentrum für Schwerionenforschung, Campus Darmstadt, 64289 Darmstadt, Germany }}
\newcommand{\EMMI}{
\affiliation{ExtreMe Matter Institute EMMI, GSI Helmholtzzentrum f\"ur Schwerionenforschung GmbH, 64291 Darmstadt, Germany}}
\newcommand{\MaxPlanckHeidelberg}{\affiliation{Max-Planck-Institut f\"ur Kernphysik, Saupfercheckweg 1, 69117 Heidelberg, Germany}}
\newcommand{\takayukinew}{\affiliation{Center for Computational Sciences, University of Tsukuba, 1-1-1 Tennodai, Tsukuba 305-8577, Japan}}

\author{M.~Enciu} 
\email{mravar@ikp.tu-darmstadt.de}
\atudarmstadt
\author{A.~Obertelli} 
\atudarmstadt \acea \ariken 
\author{P.~Doornenbal} \ariken
\author{M. Heinz} \atudarmstadt \EMMI \MaxPlanckHeidelberg
\author{T. Miyagi} \takayukinew \atudarmstadt \EMMI \MaxPlanckHeidelberg
\author{F.~Nowacki} \FNowacki
\author{K.~Ogata}\KOgataS \KOgata
\author{A.~Poves}\APoves
\author{A. Schwenk} \atudarmstadt \EMMI \MaxPlanckHeidelberg
\author{K.~Yoshida}\KYoshida
\author{N.~L.~Achouri}\acaen
\author{H.~Baba} \ariken
\author{F.~Browne}\ariken
\author{D.~Calvet} \acea
\author{F.~Ch\^ateau} \acea
\author{S.~Chen} \ahku \ariken \abeijing          
\author{N.~Chiga}\ariken 
\author{A.~Corsi} \acea 
\author{M.~L.~Cort\'es} \ariken
\author{A.~Delbart} \acea
\author{J-M.~Gheller} \acea
\author{A.~Giganon}\acea                          
\author{A.~Gillibert} \acea
\author{C.~Hilaire} \acea
\author{T.~Isobe} \ariken 
\author{T.~Kobayashi}\atohoku
\author{Y.~Kubota} \ariken \acns
\author{V.~Lapoux} \acea
\author{H.~N.~Liu} 
\atudarmstadt \acea \akth  
\author{T.~Motobayashi} \ariken                 
\author{I.~Murray} \aijclab \ariken
\author{H.~Otsu} \ariken
\author{V.~Panin}\ariken
\author{N.~Paul} \acea \abro                    
\author{W.~Rodriguez} \ariken \ajaver \aunal
\author{H.~Sakurai} \ariken \aut
\author{M.~Sasano} \ariken
\author{D.~Steppenbeck}\ariken
\author{L.~Stuhl}\acns \aatomki \aibs             
\author{Y.~L.~Sun}  \acea \atudarmstadt       
\author{Y.~Togano}\arikkyo \ariken 
\author{T.~Uesaka} \ariken
\author{K.~Wimmer}\aut \ariken                    
\author{K.~Yoneda} \ariken 
\author{O.~Aktas}\akth
\author{T.~Aumann}\atudarmstadt \agsi             
\author{L.~X.~Chung} \ainst 
\author{F.~Flavigny}\aijclab  \acaen
\author{S.~Franchoo}\aijclab 
\author{I.~Ga\v spari\'c}\azagreb \atudarmstadt \ariken
\author{R.-B.~Gerst}\akoeln
\author{J.~Gibelin}\acaen
\author{K.~I.~Hahn} \aewha \aibs
\author{D.~Kim} \aewha \ariken \aibs
\author{Y.~Kondo}\atitech                         
\author{P.~Koseoglou}\atudarmstadt \agsi
\author{J.~Lee} \ahku
\author{C.~Lehr}\atudarmstadt                     
\author{P.~J.~Li} \ahku
\author{B.~D.~Linh} \ainst 
\author{T.~Lokotko}\ahku
\author{M.~MacCormick}\aijclab
\author{K.~Moschner}\akoeln
\author{T.~Nakamura}\atitech                   
\author{S.~Y.~Park} \aewha \aibs 
\author{D.~Rossi}\atudarmstadt                  
\author{E.~Sahin} \aoslo
\author{P.-A.~S\"oderstr\"om} \atudarmstadt
\author{D.~Sohler}\aatomki  
\author{S.~Takeuchi} \atitech
\author{H.~Toernqvist}\atudarmstadt  \agsi        
\author{V.~Vaquero}\amadrid 
\author{V.~Wagner}\atudarmstadt                   
\author{S.~Wang}\alanzhou 
\author{V.~Werner}\atudarmstadt
\author{X.~Xu} \ahku
\author{H.~Yamada}\atitech                       
\author{D.~Yan} \alanzhou
\author{Z.~Yang} \ariken
\author{M.~Yasuda}\atitech                       
\author{L.~Zanetti}\atudarmstadt   

\date{\today} 
\begin{abstract}
The first spectroscopy of $^{52}$K was investigated via in-beam $\gamma$-ray spectroscopy at the RIKEN Radioactive Isotope Beam Factory after one-proton and one-neutron knockout from $^{53}$Ca and $^{53}$K beams impinging on a 15-cm liquid hydrogen target at $\approx$ 230~MeV/nucleon. The energy level scheme of $^{52}$K was built using single $\gamma$ and $\gamma$-$\gamma$ coincidence spectra. The spins and parities of the excited states were established based on momentum distributions of the fragment after the knockout reaction and based on exclusive cross sections. The results were compared to state-of-the-art shell model calculations with the SDPF-Umod interaction and ab initio IMSRG calculations with chiral effective field theory nucleon-nucleon and three-nucleon forces.\\
\\
\end{abstract}
\maketitle

\centerline{\bf I. INTRODUCTION}

Early experimental efforts in nuclear structure revealed patterns in the systematics of observables, such as binding energies or excitation energy of the first 2$^+$ states in even-even nuclei at specific numbers of protons or neutrons called \textit{magic numbers}~\cite{Meyer1948,Meyer}. 
The magic numbers for the stable nuclei are 2, 8, 20, 28, 50, 82, and 126. New ones were evidenced in unstable nuclei, see Refs.~\cite{Otsuka2020,Nowacki2021} for recent reviews. Prominent examples of new shell closures are at $N=32$ manifested in K~\cite{Rosenbusch2015}, Ca~\cite{Huck1985,Wienholtz2013,Enciu2022}, Sc~\cite{Xu2015}, Ti~\cite{Jenssens2002,Dinca2005,Leistenschneider2018}, and Cr~\cite{Prisciandaro2001,Burger2005} isotopes and at $N=34$ evidenced experimentally in Ar~\cite{Liu2018} and Ca isotopes~\cite{Michimasa2018,Chen2019,Steppenbeck2013}. The appearance of new magic numbers stems from the evolution of effective single-particle energies as a function of the proton number or the neutron number, an effect which is known as monopole drift~\cite{Talmi1960,Otsuka2020,Nowacki2021}. In the potassium isotopic chain ($Z=19$) this mechanism leads to the inversion of the $\pi$s$_{1/2}$ and $\pi$d$_{3/2}$ orbitals observed for $^{47}$K and $^{49}$K followed by the restoration of the orbitals' order in $^{51}$K and $^{53}$K~\cite{Sun}.\\
In the shell-model picture, at and near shell closures, the low-lying energy levels are of single-particle nature, while such states in mid-shell nuclei are dominated by correlations. $^{52}$K is an odd-odd nucleus surrounded by doubly-magic and semi-doubly-magic nuclei; it is one neutron away from the two exotic neutron shell closures, $N=32$ and $N=34$, and one proton away from the $Z=20$ proton shell closure. In the magic neighboring nuclei, the energy levels can be regarded as pure single-particle states~\cite{Sun,Koiwai2022,Li2024}. The energy level scheme of $^{51}$K and $^{53}$K reveal a 3/2$^-$ ground state and a 1/2$^-$ excited state created by one hole in the $\pi d_{3/2}$ and $\pi s_{1/2}$ orbitals~\cite{Sun}, respectively. On the other hand, in $^{51}_{18}$Ar, which has one proton less than $^{52}$K, one finds energy levels of mixed configuration, i.e., single-particle states and collective states~\cite{Juhasz2021}. In this paper, we present the spectroscopy of the $^{52}$K bound states populated from one proton and one neutron knockout reaction.\\

\begin{figure}
    \centering
    \includegraphics[width=0.49\linewidth]{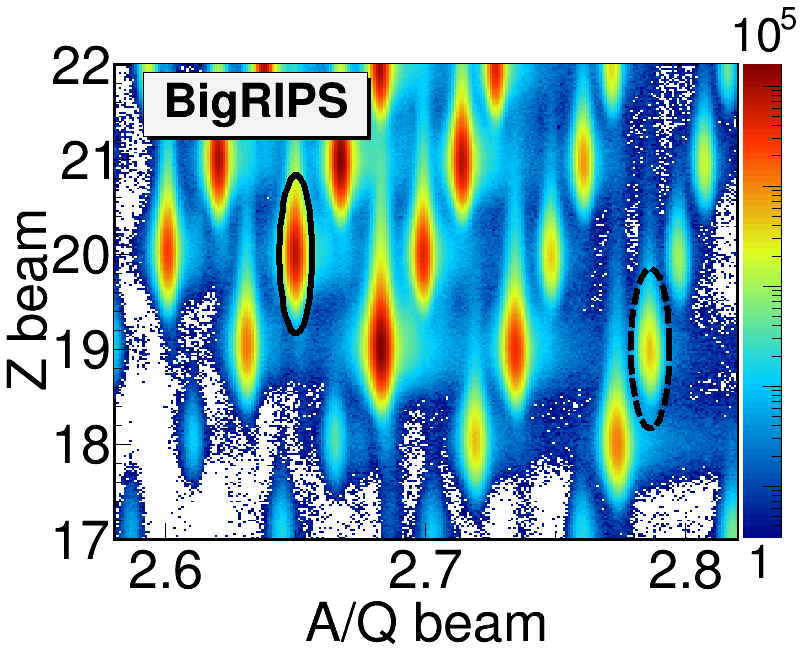}
    \includegraphics[width=0.49\linewidth]{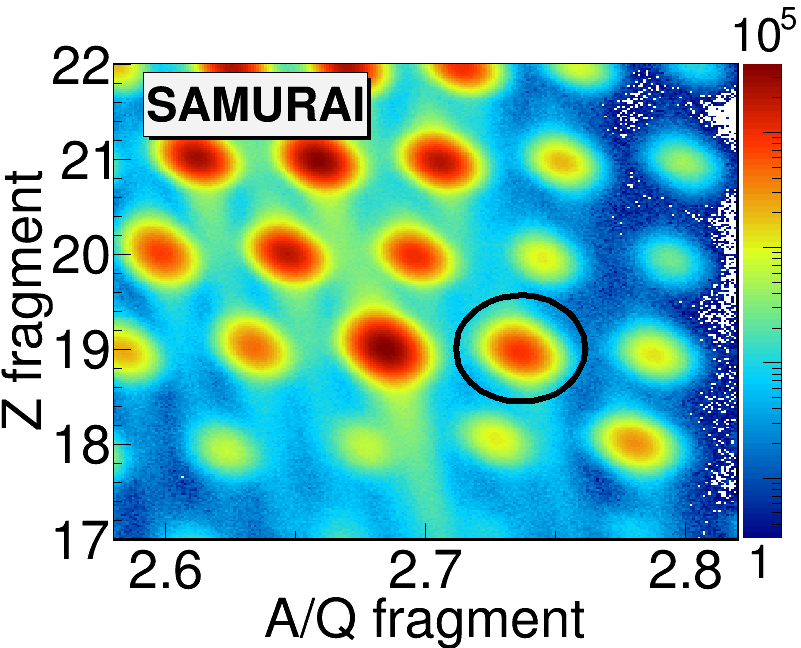}
    \caption{The particle identification (PID) for the beam in the BigRIPS spectrometer (left) and for the fragments in the SAMURAI spectrometer (right). The vertical axes show the atomic number ($Z$) and the horizontal axes show the mass-to-charge ratio ($A/Q$). The intensities are plotted with a logarithmic scale. The $^{53}$Ca and $^{53}$K beams are marked with black ellipses in the left plot (solid and dashed line, respectively) and the $^{52}$K fragment is marked with a black ellipse in the right plot.}
    \label{PID}
\end{figure}

\centerline{\bf II. EXPERIMENT}

The experiment was carried out at the Radioactive Isotope Beam Factory of RIKEN operated by the RIKEN Nishina Center and the Center for Nuclear Study of the University of Tokyo. A primary beam of $^{70}$Zn was impinged on a 10-mm thick $^{9}$Be target. The energy of the primary beam was 345~MeV/nucleon with an intensity of 240~pnA. The cocktail beam produced by fragmentation was identified event-by-event in the BigRIPS spectrometer via a B$\rho$ - ToF - $\Delta$E (magnetic rigidity - time of flight - energy loss) method~\cite{BigRIPS-T.Kubo,BigRIPS-N.Fukuda}. The beams of $^{53}$Ca and $^{53}$K had a purity of 6.5\% and 0.4\%, respectively, out of the total cocktail beam and had an intensity of 12~pps and 0.8~pps, respectively, over 7 days of beamtime. \\
The MINOS~\cite{MINOS-A.Obertelli} liquid hydrogen target was used for the proton-induced proton and neutron knockout reactions on the $^{53}$Ca and $^{53}$K beams, respectively. The target was 15.1(1)~cm long with a density of 73~mg/cm$^{3}$. The beam particles lose $\approx$100~MeV/nucleon within the target, spanning an energy range between 170 and 270 ~MeV/nucleon at the reaction vertex. The protons scattered from the reactions were tracked in the MINOS time-projection chamber (TPC) which was 30~cm long, surrounding the liquid hydrogen target. The angular coverage of the MINOS TPC was between 10$^{\circ}$ and 80$^{\circ}$ relative to the reaction vertex, with a reaction vertex reconstruction resolution of 5~mm full-width at half-maximum (FWHM)~\cite{MINOS-C.Santamaria}. The reaction vertex was constructed from the track of the two protons in the case of $(p,2p)$ reactions and from the track of the beam and one proton for the $(p,pn)$ reaction.\\
The DALI2+~\cite{DALI-S.Takeuchi}, 226-NaI(Tl)-scintillators array, was placed around the MINOS system. DALI2+ was used for detecting the $\gamma$-rays emitted by the $^{52}$K fragments after the knockout reaction. The scintillation array had a full-energy peak detection efficiency of 20\% (prior to add-back correction), an energy resolution of 10\% (FWHM) at 1~MeV, and a time resolution below 5~ns~\cite{DALI-S.Takeuchi}. DALI2+ covered the angles between 10$^{\circ}$ and 130$^{\circ}$ with respect to the reaction vertex position. The detectors were calibrated with radioactive sources ($^{137}$Cs, $^{88}$Y, and $^{60}$Co) and the energy calibration was monitored by source measurements before, after, and twice during the experiment. The energy threshold of the DALI2+ detection ranged between 50-120~keV, evaluated for each crystal individually. Simulations using the GEANT4~\cite{geant4-S.Agostinelli} framework were performed for the response function of the DALI2+ detectors. The simulated spectra were benchmarked on the radioactive source measurements. A maximum of 5\% relative difference between the experimental and the simulated spectra was measured and was considered in the systematic uncertainties for cross-section evaluation.\\
The fragments were further identified in the large-acceptance SAMURAI~\cite{SAMURAI-T.Kobayashi} spectrometer with a B$\rho$ - ToF - $\Delta$E method. The SAMURAI magnet was operated with a magnetic field of 2.9~T. The particle identification of the beam and fragments was reached with a separation of 4.7~$\sigma$ and 7.2~$\sigma$, respectively for the atomic number ($Z$) and 31.4~$\sigma$ and 8.1~$\sigma$, respectively for the mass-to-charge ratio ($A/Q$). The plots with the particle identification in BigRIPS and SAMURAI are illustrated in Fig.~\ref{PID}. Together with the particle identification the velocities of the beam and fragments were also obtained in the two spectrometers with resolutions $\frac{\Delta\beta}{\beta}$ $\approx 10^{-4}$ and $\approx 10^{-3}$, respectively. The beam and fragment selection was done using 4-$\sigma$ cuts for both $Z$ and $A/Q$. Corrections for the energy loss for the beam (fragment) particles between the BigRIPS (SAMURAI) spectrometer and the reaction vertex position were applied. The direction of the beam before the reaction vertex was obtained using the two beam drift chambers placed before MINOS and the vertex position information from MINOS. For the fragment particles, the direction was determined using the reaction vertex information and the position information in the fragment drift chamber placed after MINOS and before the SAMURAI magnet.\\
The experimental setup also included neutron detectors, NeuLAND~\cite{NeuLAND-T.Aumann} and NEBULA~\cite{NEBULA-T.Nakamura}, placed at 0$^{\circ}$ after the SAMURAI magnet at 11~m and 14~m away, respectively. NeuLAND consisted of 400 scintillator bars with the dimension of 5~cm $\times$ 5~cm $\times$ 250~cm arranged in 8 layers with alternating horizontal and vertical directions. NEBULA consisted of 120 scintillator bars of 12~cm $\times$ 12~cm $\times$ 180~cm in size, arranged in 2 walls of two layers each. The role of the neutron detection is discussed further in the text.\\

\begin{figure}
    \centering
    \includegraphics[width=0.99\linewidth]{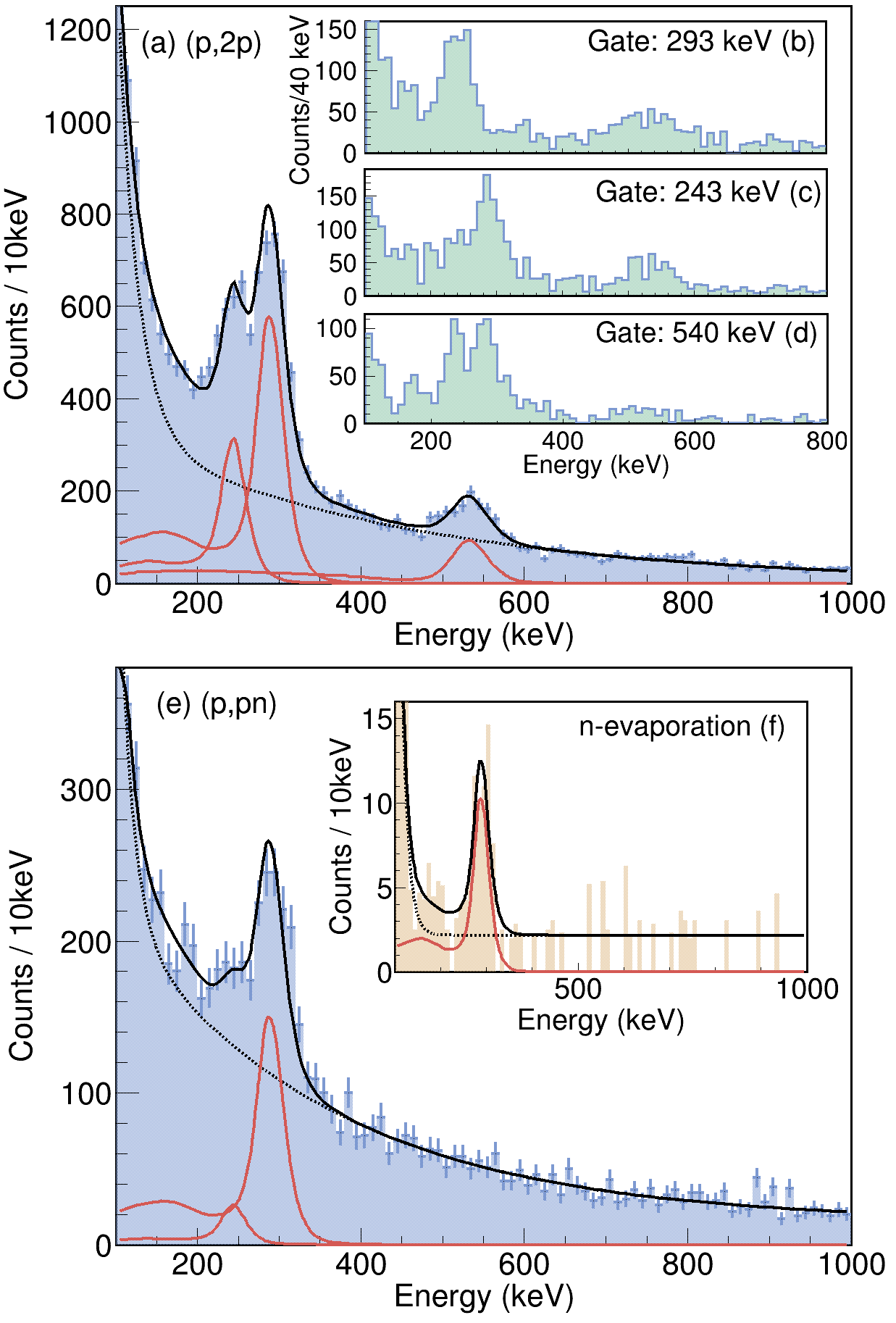}
    \caption{(a) Doppler-shift-corrected $\gamma$-ray spectrum for the $^{53}$Ca$(p,2p)^{52}$K reaction. (b-d) $\gamma$-$\gamma$ coincidence spectra with the gated transition indicated in the plot. (e) Doppler-shift-corrected $\gamma$-ray spectrum for the $^{53}$K$(p,pn)^{52}$K reaction without the neutron-evaporation subtraction. (f) $\gamma$-ray spectrum for the neutron-evaporation events corrected for the neutron detection efficiency. In the single $\gamma$-ray spectra (a, e-f), the experimental data is fitted (black line) with the simulated response functions (red line) and double-exponential decay background shape (dotted black line). }
    \label{Gamma_spec}
\end{figure}

\centerline{\bf III. METHODS AND RESULTS}

\centerline {\bf A. Gamma-ray spectroscopy}

The $(p,2p)$ reaction channel was selected by gating on the $^{53}$Ca beam, the $^{52}$K fragment, and the detection of 2 protons in MINOS which come from a reaction vertex within the target location. The $\gamma$-ray spectrum for this reaction channel corrected for the Doppler shift is shown in Fig.~\ref{Gamma_spec} (a). The spectrum is shown with a binning of 10~keV and it is fitted by the simulated response functions of DALI2+ (solid red) and a double exponential decay background shape (dotted black). The total fitting function is shown with a solid black line. The transitions identified in the spectrum have energies of 243(5)~keV, 293(3)~keV, and 540(13)~keV. For the determination of the transition energies, the simulated response functions were shifted in energy and a $\chi^2$ distribution was obtained for a fitting range around the peak energy. The transition energy was obtained from the energy at the minimum of the $\chi^2$ distribution and the uncertainties correspond to the energy between $\chi^2_{\mathrm{min}}$ and $\chi_{\mathrm{min}}^2+1$. Moreover, the $\gamma$-ray spectrum with a multiplicity condition of 1 shows mainly the 293-keV transition, indicating a direct feeding to the ground state.\\

\begin{figure}
    \centering
    \includegraphics[width=0.99\linewidth]{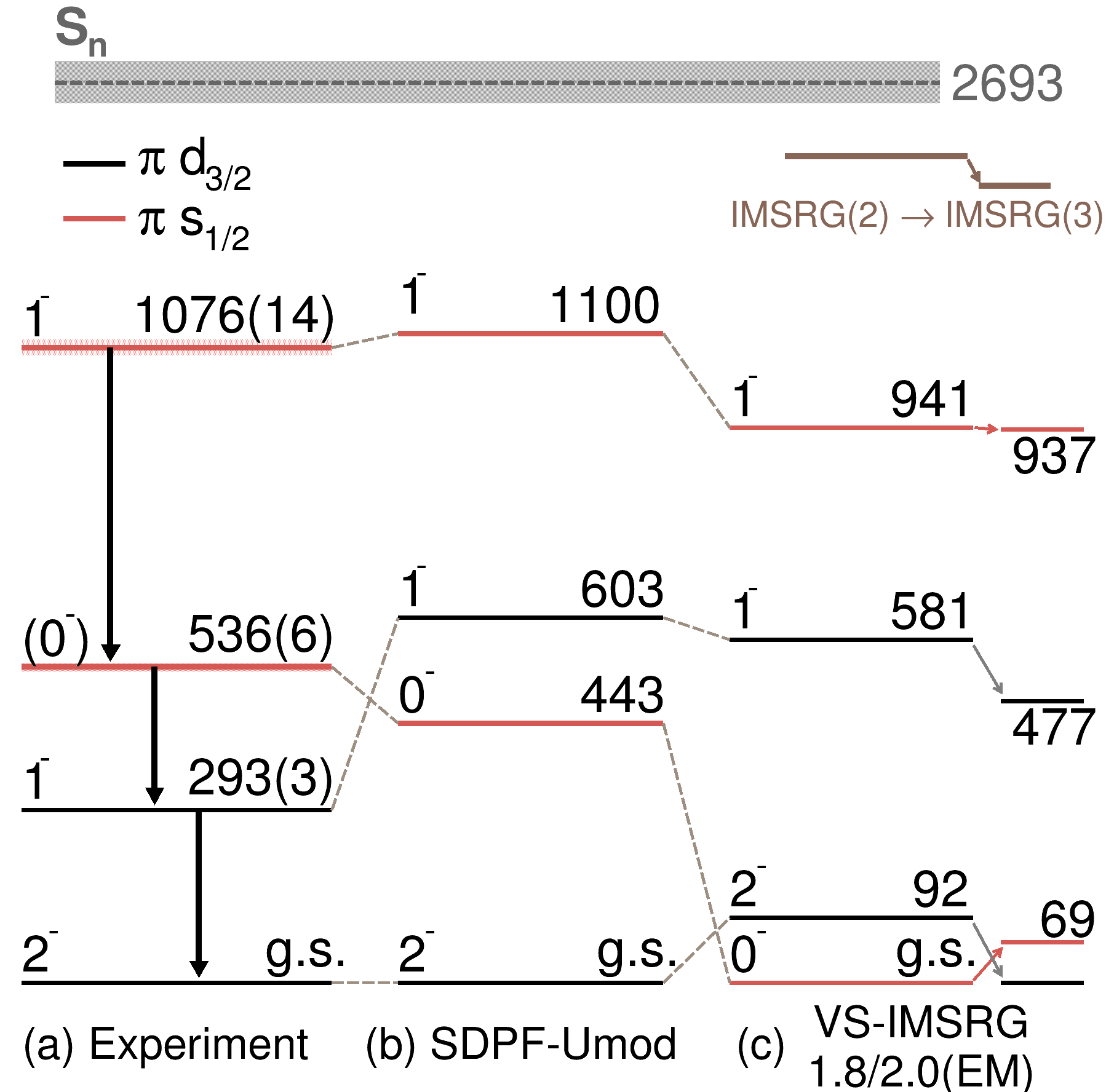}
    \caption{The experimental (a) and calculated level scheme of $^{52}$K with (b) the SDPF-Umod interaction and (c) based on the VS-IMSRG with the 1.8/2.0 (EM) interaction. See text for details about IMSRG(2) and IMSRG(3). The levels populated by s$_{1/2}$ proton knockout from $^{53}$Ca are colored in red and the levels populated by d$_{3/2}$ proton knockout are colored in black. The excitation energies are indicated and given in keV. The neutron separation energy (S$_n$) of $^{52}$K~\cite{AME2020} is marked with a grey area.}
    \label{Level_scheme}
\end{figure}

To build the energy level scheme of $^{52}$K, $\gamma$-$\gamma$ coincidence spectra were used and are shown in Fig.~\ref{Gamma_spec}, insets (b-d). Both the 243-keV and the 540-keV transitions are found in coincidence with the 293-keV transition. The coincidence spectra when gating on the 243-keV (c) and the 540-keV (d) transitions show as well that the three transitions are forming a cascade. The ratio of the 243-keV and 540-keV transitions in coincidence with the 293-keV transition implies that the 243-keV transition feeds directly the 293-keV transition. Based on these observations, we construct the energy level scheme shown in Fig.~\ref{Level_scheme}~(a) with the energy levels at 293(3), 536(6), and 1076(14) keV. \\

The $\gamma$-ray spectrum in coincidence with the $^{53}$K beam, the $^{52}$K fragment, and the detection of 1 proton in MINOS with a reaction vertex within the target region is shown in Fig.~\ref{Gamma_spec} (e) with the same color-coding as for the $(p,2p)$ spectrum. The transitions observed in this case are the 293-keV transition and 243-keV transition, the latter being significantly weaker. Because the neutron separation energy of $^{53}$K is low (S$_{n}$ = 3229(117)~keV~\cite{AME2020}), the proton inelastic scattering followed by neutron evaporation reaction can take place with the same selection condition as for $(p,pn)$. This contribution is subtracted in order to obtain a pure $(p,pn)$ reaction channel. In the $(p,pn)$ reaction, due to the kinematics, when a proton is detected in the MINOS TPC, the neutron would be scattered at similar large angles. Because there is no neutron detection at large angles around the MINOS TPC, the neutrons from the $(p,pn)$ reactions go undetected. For neutron evaporation, on the other hand, the neutrons are emitted isotropically in the center of mass which means very forward angles in the laboratory frame. We can detect the evaporated neutrons in the forward-angle NeuLAND and NEBULA neutron detectors. The detection efficiency including the geometrical acceptance and the intrinsic detector efficiency for the neutron evaporation channel was evaluated to be 39(4)\% with an energy dependence on the relative energy of \hbox{$^{52}$K + n} based on GEANT4 simulations. The $\gamma$-ray spectrum for the neutron-evaporation channel after the E$_{rel}$-dependent neutron efficiency correction is shown in Fig.~\ref{Gamma_spec} (f). \\

\centerline {\bf B. Momentum distributions}

\begin{figure}
    \centering
    \includegraphics[width=0.99\linewidth]{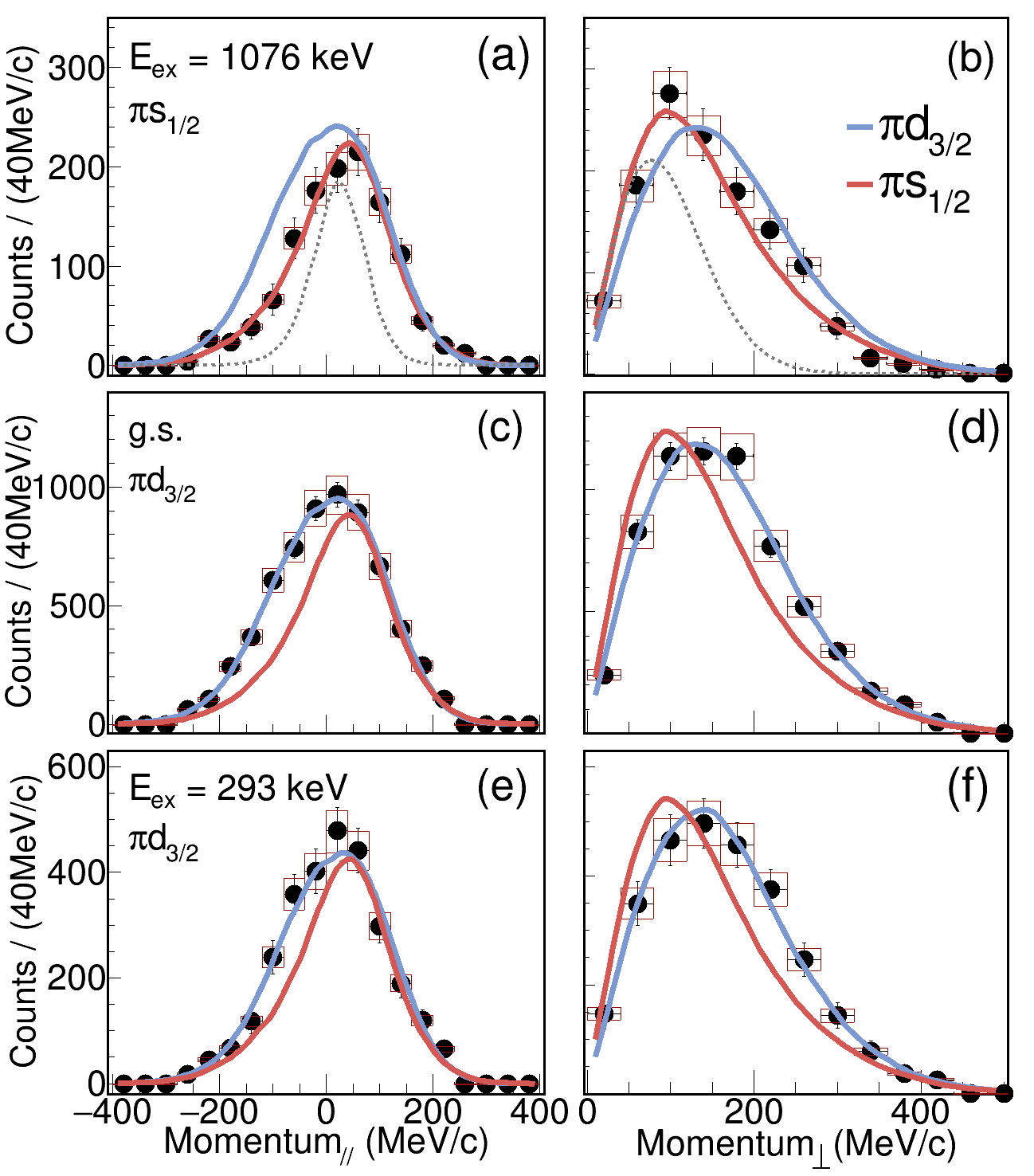}
    \caption{The momentum distributions of the $^{52}$K fragments after the proton knockout from $^{53}$Ca. The left plots show the parallel momentum component and the right plots show the perpendicular momentum component. The experimental data are plotted with black circles with statistical errors marked with bars and systematic errors marked with boxes. The experimental resolution is depicted with dashed grey lines. The DWIA theoretical curves are shown with blue for the knockout of the d$_{3/2}$ proton and with red for the s$_{1/2}$ proton. The additional d-wave in (a-b) and s-wave in (c-f) momentum distributions are overlapped with the data to highlight the difference in width.}
    \label{PMDs}
\end{figure}

The momentum distribution of the fragment relative to the beam was analyzed to constrain the spin-parity and the shell-model configuration of the excited energy levels of $^{52}$K populated by proton knockout. The momentum distributions were decomposed in the parallel and the perpendicular component with respect to the initial beam direction. The exclusive momentum distributions were obtained by fitting the corresponding $\gamma$-ray spectrum for each 40-MeV/c bin of the inclusive momentum distribution in the same manner as shown in Fig.~\ref{Gamma_spec} (a). The number of events for each populated state for a specific momentum interval is plotted separately to obtain the exclusive momentum distributions. The $\gamma$-ray spectrum fitting errors are considered for the uncertainty of the exclusive momentum distributions. The momentum distributions could be obtained only for the 293~keV and 1076~keV excited state, the statistics being too low for the population of the 536~keV excited state. The ground-state momentum distribution is obtained by the subtraction of the excited-state momentum distributions from the inclusive one. The resulting momentum distributions are shown in Fig.~\ref{PMDs} with black circles. The statistical and systematic uncertainties are plotted with lines and boxes, respectively. The resolution in momentum comes from the uncertainty in the velocity and directions of the beam and fragment. The momentum resolution was extracted from the unreacted $^{53}$Ca beam momentum distributions, plotted in Fig.~\ref{PMDs} (dotted line). A momentum resolution of 48~MeV/c was found for the parallel component and $\approx$76~MeV/c for the perpendicular components (1-$\sigma$ value). In the $(p,2p)$ reaction events, as opposed to the unreacted events, an additional degradation of momentum resolution comes from the uncertainty in the reaction vertex position. This uncertainty affects both the velocity determination at the reaction vertex and the direction determination of the beam and fragment. The uncertainty in the momentum due to the reaction vertex uncertainty is 1.7~MeV/c for the parallel component and 9.6~MeV/c for the perpendicular component.
For establishing the orbital from which the proton was knocked out to populate the ground state, the 292-keV and 1076-keV final states of $^{52}$K, the experimental data was compared to theoretical calculations for the momentum distributions.\\

\centerline {\bf C. DWIA calculations}
The single-particle cross sections and momentum distributions were calculated using the distorted-wave impulse approximation (DWIA) formalism~\cite{Chant1977,Wakasa2017,JM66,JM73,Kit85}. The calculations were performed using {\sc pikoe}~\cite{pikoe}. The optical potential for the incoming and outgoing nucleon scattering waves was constructed using the folding potential~\cite{Toyokawa2013}, employing the Melbourne $G$-matrix NN interaction~\cite{Negele2002} and the nuclear densities from HFB calculations~\cite{HFBRADcite} with the SKM interaction~\cite{SKMcite}. The NN effective interaction parametrized by Franey and Love~\cite{Franey1985} was used for describing the elementary p-n scattering process and the M{\o}ller factor~\cite{Wakasa2017} was introduced for the Lorentz transformation of the p-n scattering. The scattering and the bound-state wavefunctions were corrected for the nonlocality using the Perey factor~\cite{Per63}. The wavefunction of the knocked-out proton was calculated as a bound state of the Woods-Saxon potential. The depth of the potential was adjusted for the effective proton separation energy, the diffuseness parameter was fixed at a value of 0.67~fm, and the radial parameter was determined using the experimental momentum distributions. This approach was already used and demonstrated in Refs.~\cite{Enciu2022,Yoshida2021}. The diffuseness parameter value of 0.67~fm used in this study is based on the Bohr-Mottelson parametrization~\cite{Bohr-Mottelson}. A further study on varying both the diffuseness and the radial parameters shows that the one-dimensional variation suffices and does not impact the results~\cite{Enciu2025}. For finding the optimum radial parameter for the wavefunction of each knocked-out proton we apply the same method described in Ref.~\cite{Enciu2022}. The optimum radial parameter was found by fitting a set of theoretical curves for the momentum distributions calculated for a set of radial parameters between 0.75~fm and 1.8~fm to the experimental data. Based on the $\chi^2$ of the fit, probability distributions were built for the radial parameter from which the optimum value and the 1-$\sigma$ uncertainty are extracted. The optimum radial parameter for the $\pi d_{3/2}$ orbital using the ground state momentum distribution was found to be $r_0$ = 1.549(66)~fm, while for the $\pi s_{1/2}$ orbital using the 1076-keV excited state momentum distribution was found to be $r_0$ = 1.416(160)~fm. The root-mean-square radii of the single-particle proton orbitals extracted using this method are beyond the scope of this work and will be discussed in a separate publication~\cite{Enciu2025}. \\
The theoretical momentum curves were convoluted with the reaction energy profile and with the experimental momentum resolution. The theoretical curves (with the optimum $r_0$ parameter) are overlapped with the experimental data in Fig.~\ref{PMDs} and are marked with red for s$_{1/2}$ and with blue for the d$_{3/2}$ proton knockout. The momentum distributions for s-wave and d-wave partners which do not reproduce the experimental data are also plotted to highlight the difference in width.\\
Based on the comparison to theoretical momentum distributions, the ground state and the 293-keV excited state are populated via d$_{3/2}$ proton removal, while the 1076-keV excited state of $^{52}$K is populated by s$_{1/2}$ proton removal; this is indicated by different colors in Fig.~\ref{Level_scheme}.\\

\centerline {\bf D. Shell model and ab initio calculations}

We compare the measured spectrum of $^{52}$K with the predictions from two theoretical calculations; one is a shell-model calculation with an effective interaction, and the other is an ab initio calculation based on two- and three-nucleon forces.\\
The effective shell model spectrum is computed using the SDPF-Umod Hamiltonian as in Ref.~\cite{Sun} diagonalized in the $sd-pf$ valence space for both protons and neutrons and an $^{16}$O inert core. The calculated energy levels corresponding to the observed ones are illustrated in Fig.~\ref{Level_scheme} and their occupation numbers are listed in Table~\ref{tab_th}. One finds a good agreement between the effective shell model calculations and the experimental level scheme for $^{52}$K. 
The SDPF-Umod interaction reproduces the E($3/2^+-1/2^+$) exactly for $^{51}$K~\cite{Sun}, while this calculated quantity is slightly too small for $^{53}$K by 300~keV~\cite{Sun}. By extrapolation, we expect that the d$_{3/2}-$s$_{1/2}$ single-particle orbitals effective energy splitting is underestimated by the SDPF-Umod interaction by 150~keV. This is one of the reasons why the calculated 0$^-$ state of $^{52}$K is lower than the experimental value. At the same time, the inversion of the calculated 0$^-$ and 1$^-$ states as well as the energy splitting between the pair of states with the same occupation number configurations, namely 2$^--1^-$ and 0$^--1^-_2$, stem from the multipole recoupling of a d$_{3/2}$ or an s$_{1/2}$ proton to a p$_{1/2}$ neutron.
Additionally, the spectroscopic factors (C$^2$S) for knocking out a proton from the $^{53}$Ca ground state were calculated and the following values were found: C$^2$S = 2.352 and 1.303 for knocking out a d$_{3/2}$ proton and populating the 2$^-$ and the 1$^-$ state in $^{52}$K, respectively, and C$^2$S =  0.402 and 1.244 for knocking out an s$_{1/2}$ proton and populating the $0^-$ and 1$^-_2$ states of $^{52}$K, respectively.\\

\begin{table}[t]
\centering
  \caption{The energy levels (E$_{\mathrm{ex}}$) of $^{52}$K and the occupation numbers for the neutron ($\nu$) and proton ($\pi$) orbitals in the valence space are listed in the table for the two different theoretical models. The odd occupation numbers are marked in bold to easily guide the eye to the 1p-1h states and to identify the first 4 states of $^{52}$K observed experimentally.}
  \begin{spacing}{1.2}
\begin{tabular}{  p{1.2cm} p{0.5cm} |  p{0.8cm} p{0.8cm} p{0.8cm} p{0.8cm}   p{0.8cm} p{0.8cm} p{0.8cm}  } 
\hline
\hline
\multicolumn{2}{c |}{Energy levels}& \multicolumn{7}{c}{Occupation numbers}\\
$E_{\mathbf{ex}}$\,&$J^{\pi}$& $\nu f_{7/2}$ & $\nu p_{3/2}$ & $\nu p_{1/2}$& $\nu f_{5/2}$ & $\pi d_{5/2}$& $\pi s_{1/2}$& $\pi$d$_{3/2}$ \\
\hline
\multicolumn{9}{c}{\bf Shell Model with SDPF-Umod}\\
\hline
0     	& 2$^{-}$    & 7.9 & 3.9 & {\bf 1.0} &  0.2 & 6.0 & 2.0 & {\bf 3.0}\\
443    	& 0$^{-}$    & 7.9 & 3.8 & {\bf 1.2} & 0.2 & 6.0 & {\bf 1.2} & 3.8 \\
603  	& 1$^{-}$  	 & 7.9 & 3.9 & {\bf 1.1} & 0.2 & 6.0 & 1.9 & {\bf 3.1}\\
1100    & 1$^{-}_{2}$  & 7.9 & 3.9 & {\bf 1.0} & 0.3 & 6.0 & {\bf 1.2} & 3.9  \\
\hline
\multicolumn{9}{c}{\bf VS-IMSRG(2) with 1.8/2.0 (EM)}\\
\hline
0     	& 0$^{-}$    & 7.8 & 3.7 & {\bf 1.2} & 0.3 & 6.0 & {\bf 1.2} & 3.8  \\
92  	& 2$^{-}$  	 & 7.8 & 3.9 & {\bf 1.0} &  0.3 & 6.0 & 2.0 & {\bf 3.1}  \\
581    	& 1$^{-}$    & 7.8 & 3.8 & {\bf 1.1} & 0.3 & 6.0 & 1.8 & {\bf 3.3} \\ 
941    & 1$^{-}_{2}$  & 7.8 & 3.8 & {\bf 1.0} & 0.3 & 6.0 & {\bf 1.3} & 3.7  \\
\hline
\hline
\end{tabular}
\end{spacing}
  \label{tab_th}
\end{table}

The spectrum of $^{52}$K was also computed using the ab initio valence-space in-medium similarity renormalization group (VS-IMSRG)~\cite{Tsukiyama2011,Hergert2016,Stroberg2017,Stroberg2019}. The VS-IMSRG aims to solve the many-body Schrödinger equation starting from inter-nucleon interactions by computing a unitary transformation $U = e^\Omega$ to block-diagonalize the Hamiltonian. This transformation decouples a valence space from the remaining states in the Hilbert space, producing an effective valence-space Hamiltonian that can be diagonalized using shell model methods. VS-IMSRG calculations are typically truncated at the normal-ordered two-body level, the VS-IMSRG(2), but recent work has extended this truncation to the normal-ordered three-body level with VS-IMSRG(3)-$N^7$ calculations becoming available~\cite{Heinz2021,Stroberg2024}. 
The calculations with the VS-IMSRG(2) start from a Hartree-Fock reference state, decoupling a valence space with an $^{28}$O core and proton $sd$ and neutron $fp$ valence orbitals. The 1.8/2.0~(EM) Hamiltonian~\cite{Hebeler2012} is used, which has been shown to give accurate spectra in medium-mass nuclei~\cite{Hagen2016,Taniuchi2019,Simonis2017,Stroberg2019}. The calculations are performed in a model space of 15 major harmonic oscillator shells ($e = 2n + l \leq e_\text{max} = 14$) with an underlying frequency $\hbar\omega= 16\:\text{MeV}$. For three-nucleon forces, an additional three-body truncation is imposed on three-body states $\ket{pqr}$ such that $e_p + e_q + e_r \leq E_\text{3max} = 24$. The \textsc{imsrg++} code~\cite{StrobergIMSRG} is used for the VS-IMSRG calculations and the \textsc{kshell} code~\cite{ShimizuKSHELL} is used for the final shell model diagonalization. 
The obtained energy levels and the occupation numbers are listed in Table~\ref{tab_th}. A small difference of 92~keV is found between the predicted energies of the $0^-$ and $2^-$ states with a $0^-$ ground state.\\

To understand the uncertainties of the VS-IMSRG(2) truncation, we perform VS-IMSRG(3)-$N^7$ calculations. \\
The induced three-body operators included in the VS-IMSRG(3) are computationally expensive, so we employ a restricted model space: we include three-body matrix elements with single-particle states with $2n+l \leq e_\mathrm{max,3b} = 4$, including all states with $E_\text{3max} = 3 e_\mathrm{max,3b} = 12$~\cite{Heinz2024}. We emphasize that the model space used is insufficient to fully converge the effects of three-body operators (which would require taking $e_\mathrm{max,3b} = 14$), but serves to give insight into the sizes of VS-IMSRG(3) corrections to the predicted spectrum. We find that the level structure is affected by the many-body truncation. Our unconverged VS-IMSRG(3)-$N^7$ calculations suggest a strong preference for a 2$^-$ ground state as in the experiment and a 69-keV excitation energy for the 0$^-$ state. The energy of the 0$^-$ state is expected to further increase for fully converged VS-IMRSG(3) calculations. The evolution of the energy levels from VS-IMSRG(2) to VS-IMSRG(3)-$N^7$ is shown in Fig.~\ref{Level_scheme}.\\ 

\begin{table*}[t]
\centering
  \caption{The table lists the experimental energy levels (E$_{\mathrm{ex}}^{\mathrm{exp}}$) and the assigned spin-parities (J$^{\pi}$) in the first two columns. The next two columns list the experimental inclusive and exclusive cross sections ($\sigma_{-1p}^{\mathrm{exp}}$) for the $^{53}$Ca$(p,2p)^{52}$K reaction and the orbital from which the proton was knocked out is specified. For each final state, the calculated single-particle cross section ($\sigma^{\mathrm{DWIA}}_{\mathrm{sp},-1p}$) is given together with the spectroscopic factors (C$^2$S) calculated with the effective interaction SDPF-Umod. The theoretical cross sections, which are obtained as the product of $\sigma^{\mathrm{DWIA}}_{\mathrm{sp},-1p}$ and C$^2$S, are listed under $\sigma_{-1p}^{\mathrm{th}}$. The calculations using the optimum radial parameter and the default Bohr-Mottelson parametrization are labeled with (a) and (b), respectively. See text for details. In the last three columns, the relative cross sections, $\sigma_{\mathrm{E}_{\mathrm{ex}}}/\sigma_{\mathrm{inclusive}}$ are listed. The relative cross sections for the experimental values and for theoretical values with methods (a) and (b) are given for each energy level of $^{52}$K in units of percentages (\%). The energies are given in keV and the cross sections are given in mb.}
  \begin{spacing}{1.2}
\begin{tabular}{ p{1.8cm} p{1cm} | p{1cm} p{1.8cm}   p{1.cm} p{1.cm}  p{1.cm} p{1.cm} p{1.cm} | p{1.8cm} p{1.5cm} p{1.5cm}} 
\hline
\hline
E$_{\mathrm{ex}}^{\mathrm{exp}}$\,&$J^{\pi}$&$-1p$&\,\,\,$\sigma_{-1p}^{\mathrm{exp}}$&   \multicolumn{2}{l}{\,\,\,$\sigma_{\mathrm{sp},-1p}^{\mathrm{DWIA}}$}    &  \multicolumn{1}{l}{C$^2$S$^{\mathrm{SM}}$} & \multicolumn{2}{l|}{\,\,\,\,\,$\sigma^{\mathrm{th}}_{-1p}$}& \multicolumn{3}{c}{Ratio $\sigma_{\mathrm{E}_{\mathrm{ex}}}/\sigma_{\mathrm{inclusive}}$}\\
\multicolumn{2}{c|}{ }& \multicolumn{2}{c}{ }& (a) & (b) & \multicolumn{1}{c}{ }  & (a) & (b) &\,\,\,$\sigma_{-1p}^{\mathrm{exp}}$&$\sigma^{\mathrm{th}}_{-1p}$ (a)&$\sigma^{\mathrm{th}}_{-1p}$ (b)\\
\hline
0     		& 2$^{-}$    	& d$_{3/2}$   	& 3.21(43)		& 1.67 & 1.01	& 2.35 & 4.23& 2.56	& \,\,55(8)\%& \,\,47\% & \,\,44\%\\
293(3)  	& 1$^{-}$  		& d$_{3/2}$    	& 1.45(21)		& 1.67 & 1.01	& 1.30 & 2.18& 1.32	& \,\,25(4)\% & \,\,24\% & \,\,23\%\\
536(6)    	& (0$^{-})$   	& s$_{1/2}$   	& 0.38(10)		& 1.60 & 1.17	& 0.40 & 0.64& 0.47	& \,\,7(2)\% & \,\,7\% & \,\,8\%\\
1076(14)    & 1$^{-}$     	& s$_{1/2}$    	& 0.81(12)		& 1.60 & 1.17	& 1.24 & 1.99& 1.46	& \,\,14(3)\% & \,\,22\% & \,\,25\%\\
\hline
inclusive	&\multicolumn{1}{c|}{ }&	\multicolumn{1}{c}{ }& 5.84(39)	&\multicolumn{3}{c}{ } & 9.04 & 5.80	&		\\
\hline
\hline
\end{tabular}
\end{spacing}
  \label{tab_exp}
\end{table*}

The first 4 levels calculated from the effective shell model and the ab initio VS-IMSRG Hamiltonian are shown in Fig.~\ref{Level_scheme} together with the experimental energy level scheme. In all calculations, we observe two energy levels (2$^-$ and 1$^-$) obtained by coupling a p$_{1/2}$ neutron with a d$_{3/2}$ proton and two energy levels (0$^-$ and 1$^-$) obtained by coupling a p$_{1/2}$ neutron with an s$_{1/2}$ proton. There is a good agreement (within 0.5~MeV) between the experimental energies of the levels and the calculated ones. Guided by the calculated energy levels and the exclusive momentum distributions, the ground state is assigned J$^{\pi}$=2$^{-}$ and the 293-keV state found experimentally is assigned J$^{\pi}$=1$^{-}$. Similarly, the experimentally found 1076-keV excited state is assigned the spin-parity of 1$^{-}_2$. The momentum distribution for the 536-keV final state could not be analyzed, but based on the analogy with the theoretical calculations, it is tentatively assigned J$^{\pi}$=0$^{-}$. The calculated occupation numbers are very similar for the two different interactions and show the single-particle nature of the states in $^{52}$K. To quantify the single-particle nature of the energy levels experimentally we compare the experimental and theoretical exclusive cross sections.\\

\centerline {\bf E. Exclusive cross sections}

The experimental inclusive and exclusive cross-sections were evaluated for both the $(p,2p)$ and $(p,pn)$ reaction channels to further confirm the energy level scheme built for $^{52}$K and to quantify the single-particle nature of the states. 
An inclusive cross section of 5.84(39)~mb was obtained for the $(p,2p)$ reaction with  3.21(43)~mb, 1.45(21)~mb, 0.38(10)~mb, and 0.81(12)~mb for the 2$^-$, 1$^-$, 0$^-$, and 1$^-_2$ state, respectively, listed in Table~\ref{tab_exp}. For the $(p,pn)$ reaction, an inclusive cross section of 30.6(23)~mb was determined with 20.6(31)~mb, 8.5(16)~mb, and 1.54(66)~mb for the  2$^-$, 1$^-$, and 0$^-$ state, respectively; these values are obtained after the subtraction of the neutron evaporation contribution which amounts to $\approx$10\% out of the $(p,pn)$ events and has an inclusive cross section of 3.28(43)~mb. The uncertainties given for the cross-sections contain both statistical and systematic uncertainties. For the systematic uncertainties, the following were considered: spectrum fitting systematic uncertainty (10\%); DALI2+ efficiency systematic uncertainty (5\%), MINOS efficiency evaluation systematic uncertainty (6\%), and finally, the systematic uncertainty due to the neutron detection efficiency evaluation (10\%) in the case of the neutron-evaporation cross-section subtraction only.  \\

Using the optimum $r_0$ values (a) for the single-particle cross-section calculation with the DWIA, one obtains 1.67~mb and 1.60~mb for the $\pi d_{3/2}$ and $\pi s_{1/2}$ orbitals, respectively. If one uses the Bohr-Mottelson parametrization~\cite{Bohr-Mottelson} with $r_0$=1.27~fm (b), the single-particle cross sections become 1.01~mb and 1.17~mb, respectively. The theoretical cross sections are calculated as the product of the single-particle cross sections and the spectroscopic factors from the effective shell model, $\sigma^{th} = \sigma_{sp}^{DWIA} \cdot C^2S$.  The theoretical inclusive cross section using the optimum radial parameter is 9.04~mb and for $r_0$=1.27~fm is 5.80~mb. The cross sections calculated for both the optimum $r_0$ and $r_0$=1.27~fm are listed in Table~\ref{tab_exp}. We obtain a quenching factor ($\sigma^{exp}/\sigma^{th}$) of 0.66 for the $(p,2p)$ inclusive cross section by using the optimum radial parameter. This finding agrees with the quenching factors for $(e,e'p)$ knockout reactions on stable nuclei~\cite{Aumann2021}. No quenching factor (R$_{s}$=1.01) is found when the Bohr-Mottelson radial parameter is used as in Ref.~\cite{Sun}. \\

The exclusive relative cross sections ($\sigma_{\mathrm{E}_{\mathrm{ex}}}/\sigma_{\mathrm{inclusive}}$) from the experiment and theory, (a) and (b), were compared and are listed in the last three columns of Table~\ref{tab_exp}. One finds about half of the cross section for the ground state, a quarter for the $1^-$ state, up to 10\% for the $0^-$ state, and up to 25\% for the $1^-_2$ state. There is an excellent agreement of both (a) and (b) calculations with the experimental relative exclusive cross sections; the largest deviation is found for the $1^-_2$ state, which in turn affects the ground state values.\\ 

The comparison of the $(p,2p)$ cross sections to the theoretical ones further confirms the spin-parity assignment of the energy levels of $^{52}$K and the single-particle nature of the energy levels. The $(p,pn)$ cross sections show large values for the ground state, mainly, and the 293-keV state, for which the proton orbital occupation of $^{53}$Ca$_{g.s.}$ and $^{52}$K is similar, while little or zero cross section for other states with different proton configuration, as expected for the presented level scheme of $^{52}$K.\\

\centerline{\bf IV. CONCLUSIONS}

The spectroscopy of $^{52}$K was investigated via $\gamma$-ray spectroscopy after one-proton $(p,2p)$ and one-neutron $(p,pn)$ knockout reactions at $\approx$~230~MeV/nucleon. Three transitions were observed in the single $\gamma$-ray and $\gamma$-$\gamma$ coincidence spectra. It is shown that $^{52}$K has a 2$^-$ ground state and three excited states at 293(3)~keV ($1^-$), 536(6)~keV ($0^-$), and 1076(14)~keV ($1^-_2$). For the spin-parity assignment, the experimental exclusive momentum distributions were analyzed and compared to DWIA calculations. Calculations for the energy levels scheme were performed with two models, the shell model with the SDPF-Umod interaction and ab initio VS-IMSRG(2) with the 1.8/2.0~(EM) interaction. The calculated energies are in agreement (within 0.5~MeV) with the experimental level scheme and guided the spin-party assignment. Additionally, the energy level scheme and the single-particle nature of the states are further confirmed by the agreement between the experimental and theoretical relative exclusive cross sections. The energy levels of $^{52}$K exhibit a single-particle nature, dominated by proton excitations of low energy. In $^{52}$K, the four energy levels presented in this work arise from one-particle one-hole configurations, coupling a p$_{1/2}$ neutron to a d$_{3/2}$ proton or a s$_{1/2}$ proton respectively, with the two possible spin orientations. Moreover, we find in this study that the s$_{1/2}$ and d$_{3/2}$ proton orbitals keep the normal ordering in $^{52}$K, with an energy gap similar to its neighbors $^{51,53}$K~\cite{Sun}.\\

\begin{acknowledgments}
We are grateful for the support of the RIKEN Nishina Center accelerator staff in the delivery of the primary beam and the BigRIPS team for preparing the secondary beams. The development of MINOS has been supported by the European Research Council through the ERC Grant No. MINOS-258567. M. E., A. O., A. S., T. A., I. G., C. L., D. R., H. T., V. W., and L. Z. acknowledge the support from the Deutsche Forschungsgemeinschaft (DFG, German Research Foundation) -- Project-ID. 279384907 -- SFB 1245. K. O. acknowledges the support by Grants-in-Aid for Scientific Research from the JSPS (No. JP21H00125). M. H., T. M., and A.S.~were supported in part by the European Research Council (ERC) under the European Union's Horizon 2020 research and innovation programme (Grant Agreement No.~101020842). T. M. is supported by JST ERATO Grant No. JPMJER2304, Japan. A. P. is funded by Grant CEX2020-001007-S funded by MCIN/AEI/10.13039/501100011033 and PID2021-127890NB-I00. Y. T. acknowledges the support from the JSPS Grant-in-Aid for Scientific Research Grants No. JP21H01114. B. D. L. and L. X. C. acknowledge support from the Vietnam Ministry of Science and Technology under Grant No. {\DJ}TCB.01/21/VKHKTHN. D. S. acknowledges the National Research, Development and Innovation Fund of Hungary via project No. K128947. F. B. was supported by the RIKEN Special Postdoctoral Researcher Program. V. W. acknowledges the BMBF Grant Nos. 05P21RDFN1 and 05P21RDFN9.\par
\end{acknowledgments}


\begin{thebibliography}{99}
\bibitem{Meyer1948} M. G. Mayer, “On closed shells in nuclei”, Phys. Rev. {\bf 74}, 235–239 (1948).
\bibitem{Meyer} M. Mayer and J. H. D. Jensen, {\it Elementary Theory of Nuclear Shell Structure} (Wiley, New York, 1955).
\bibitem{Otsuka2020} T. Otsuka, A. Gade, O. Sorlin, T. Suzuki, and Y. Utsuno, Rev. Mod. Phys. {\bf 92}, 015002 (2020).
\bibitem{Nowacki2021} F. Nowacki, A. Poves, and A. Obertelli, Prog. Part. Nucl. Phys. {\bf 120}, 103866 (2021).
\bibitem{Rosenbusch2015} M. Rosenbusch, P. Ascher, D. Atanasov, C. Barbieri, D. Beck, K. Blaum, Ch. Borgmann, M. Breitenfeldt, R. B. Cakirli, A. Cipollone {\it et al.}, Phys. Rev. Lett. {\bf 114}, 202501 (2015). 
\bibitem{Huck1985} A. Huck, G. Klotz, A. Knipper, C. Miehe, C. Richard-Serre, G. Walter, A. Poves, H. L. Ravn, G. Marguier {\it et al.}, Phys. Rev. C {\bf 31}, 2226 (1985).
\bibitem{Wienholtz2013} F. Wienholtz, D. Beck, K. Blaum, Ch. Borgmann, M. Breitenfeldt, R. B. Cakirli, S. George, F. Herfurth, J. D. Holt, M. Kowalska {\it et al.}, Nature {\bf 498}, 346–349 (2013).
\bibitem{Enciu2022} M. Enciu, H. N. Liu, A. Obertelli, P. Doornenbal, F. Nowacki, K. Ogata, A. Poves, K. Yoshida, N. L. Achouri, H. Baba {\it et al.}, Phys. Rev. Lett. {\bf 129}, 262501 (2022).
\bibitem{Xu2015} X. Xu, M. Wang, Y.-H. Zhang, H.-S. Xu, P. Shuai, X.-L. Tu, Y. A. Litvinov, X.-H. Zhou, B.-H. Sun, Y.-J. Yuan {\it et al.}, Chin. Phys. C {\bf 39}, 104001 (2015).
\bibitem{Jenssens2002} R. V. F Janssens, B. Fornal, P. F. Mantica, B. A. Brown, R. Broda, P. Bhattacharyya, M. P. Carpenter, M. Cinausero, P. J. Daly, A. D. Davies {\it et al.}, Phys. Lett. B {\bf 546}, 55-62 (2002).
\bibitem{Dinca2005} D.-C. Dinca, R. V. F. Janssens, A. Gade, D. Bazin, R. Broda, B. A. Brown, C. M. Campbell, M. P. Carpenter, P. Chowdhury, J. M. Cook {\it et al.}, Phys. Rev. C {\bf 71}, 041302 (2005).
\bibitem{Leistenschneider2018} E. Leistenschneider, M. P. Reiter, S. Ayet San Andrés, B. Kootte, J. D. Holt, P. Navrátil, C. Babcock, C. Barbieri, B. R. Barquest, J. Bergmann {\it et al.}, Phys. Rev. Lett. {\bf 120}, 062503 (2018).
\bibitem{Prisciandaro2001} J. Prisciandaro, P. Mantica, B. Brown, D. Anthony, M. Cooper, A. Garcia, D. Groh, A. Komives, W. Kumarasiri, P. Lofy {\it et al.}, Phys. Lett. B {\bf 510}, 17–23 (2001).
\bibitem{Burger2005} A. B\"urger, T. Saito, H. Grawe, H. H\"ubel, P. Reiter, J. Gerl, M. G\'orska, H. Wollersheim, A. Al-Khatib, A. Banu {\it et al.}, Phys. Lett. B {\bf 622}, 29–34 (2005).
\bibitem{Liu2018} H. N. Liu, A. Obertelli, P. Doornenbal, C. A. Bertulani, G. Hagen, J. D. Holt, G. R. Jansen, T. D. Morris, A. Schwenk, R. Stroberg, {\it et al.}, Phys. Rev. Lett. {\bf 122}, 072502 (2019).
\bibitem{Michimasa2018} S. Michimasa, M. Kobayashi, Y. Kiyokawa, S. Ota, D. S. Ahn, H. Baba, G. P. A. Berg, M. Dozono, N. Fukuda, T. Furuno {\it et al.}, Rev. Lett. {\bf 121}, 022506 (2018).
\bibitem{Chen2019} S. Chen, J. Lee, P. Doornenbal, A. Obertelli, C. Barbieri, Y. Chazono, P. Navr\'atil, K. Ogata, T. Otsuka, F. Raimondi {\it et al.}, Phys. Rev. Lett. {\bf 123}, 142501 (2019).
\bibitem{Steppenbeck2013} D. Steppenbeck, S. Takeuchi, N. Aoi, P. Doornenbal, M. Matsushita, H. Wang, H. Baba, N. Fukuda, S. Go, M. Honma {\it et al.}, Nature {\bf 502}, 207–210 (2013).
\bibitem{Talmi1960} I. Talmi and I. Unna, Phys. Rev. Lett. {\bf 4}, 469 (1960).
\bibitem{Sun} Y. L. Sun, A. Obertelli, P. Doornenbal, C. Barbieri, Y. Chazono, T. Duguet, H. N. Liu, P. Navr\'atil, F. Nowacki, K. Ogata {\it et al.}, Phys. Lett. B {\bf 802}, 135215 (2020).
\bibitem{Koiwai2022} T. Koiwai, K. Wimmer, P. Doornenbal, A. Obertelli, C. Barbieri, T. Duguet, J. D. Holt, T. Miyagi, P. Navr\'atil, K. Ogata {\it et al.}, Phys. Lett. B {\bf 827}, 136953 (2022).
\bibitem{Li2024} P. J. Li, J. Lee, P. Doornenbal, S. Chen, S. Wang, A. Obertelli, Y. Chazono, J. D. Holt, B. S. Hu, K. Ogata {\it et al.}, Phys. Lett. B {\bf 855} 138828 (2024).
\bibitem{Juhasz2021} M. M. Juhász, Z. Elekes, D. Sohler, Y. Utsuno, K. Yoshida, T. Otsuka, K. Ogata, P. Doornenbal, A. Obertelli, H. Baba {\it et al.}, Phys. Lett. B {\bf 814}, 136108 (2021).
\bibitem{BigRIPS-T.Kubo} T. Kubo, D. Kameda, H. Suzuki, N. Fukuda, H. Takeda, Y. Yanagisawa, M. Ohtake, K. Kusaka, K. Yoshida, N. Inabe {\it et al.}, Prog. Theor. Exp. Phys. {\bf 2012}, 03C003 (2012).
\bibitem{BigRIPS-N.Fukuda} N. Fukuda, T. Kubo, T. Ohnishi, N. Inabe, H. Takeda, D. Kameda, and H. Suzuki, Nucl. Instrum. Methods Phys. Res. B {\bf 317}, 323 (2013).
\bibitem{MINOS-A.Obertelli} A. Obertelli, A. Delbart, S. Anvar, L. Audirac, G. Authelet, H. Baba, B. Bruyneel, D. Calvet, F. Ch\^ateau, A. Corsi {\it et al.}, Eur. Phys. J. A {\bf 50}, 8 (2014).
\bibitem{MINOS-C.Santamaria} C. Santamaria, A. Obertelli, S. Ota, M. Sasano, E. Takada, L. Audirac, H. Baba, D. Calvet, F. Ch\^ateau, A. Corsi {\it et al.}, Nucl. Instrum. Methods Phys. Res. A {\bf 905}, 138 (2018).
\bibitem{DALI-S.Takeuchi} S. Takeuchi, T. Motobayashi, Y. Togano, M. Matsushita, N. Aoi, K. Demichi, H. Hasegawa, and H. Murakami, Nucl. Instrum. Methods Phys. Res. A {\bf 763}, 596 (2014).
\bibitem{geant4-S.Agostinelli} S. Agostinelli, J. Allison, K. Amako, J. Apostolakis, H. Araujo, P. Arce, M. Asai, D. Axen, S. Banerjee, G. Barrand {\it et al.}, Nucl. Instrum. Methods Phys. Res. A {\bf 506}, 250 (2003).
\bibitem{SAMURAI-T.Kobayashi} T. Kobayashi, N. Chiga, T. Isobe, Y. Kondo, T. Kubo, K. Kusaka, T. Motobayashi, T. Nakamura, J. Ohnishi, H. Okuno {\it et al.}, Nucl. Instrum. Methods Phys. Res. B {\bf 317}, 294 (2013).
\bibitem{NeuLAND-T.Aumann} K. Boretzky, I. Ga\v spari\'c, M. Heil, J. Mayer, A. Heinz, C. Caesar, D. Kresan, H. Simon, H. T. T\"ornqvist, D. K\"orper {\it et al.}, Nucl. Instrum. Methods Phys. Res. A {\bf 1014}, 165701 (2021).
\bibitem{NEBULA-T.Nakamura} T. Nakamura, and Y. Kondo, Nucl. Instrum. Methods Phys. Res. B {\bf 376}, 156 (2016).
\bibitem{AME2020} M. Wang, W. J. Huang, F. G. Kondev, G. Audi, and S. Naimi, Chin. Phys. C {\bf 45(3)}, 030003 (2021).
\bibitem{Chant1977} N. S. Chant and P. G. Roos, Phys. Rev. C {\bf15}, 57 (1977).
\bibitem{Wakasa2017} T. Wakasa, K. Ogata, and T. Noro, Prog. Part. Nucl. Phys. {\bf96}, 32 (2017).
\bibitem{JM66} G. Jacob and Th. A. J. Maris, Rev. Mod. Phys. {\bf38}, 121 (1966).
\bibitem{JM73} G. Jacob and Th. A. J. Maris, Rev. Mod. Phys. {\bf45}, 6 (1973).
\bibitem{Kit85} P. Kitching, W. J. McDonald, Th. A. J. Maris, and C. A. Z. Vasconcellos, Adv. Nucl. Phys. {\bf15}, 43 (1985).
\bibitem{pikoe} K. Ogata, K. Yoshida, and Y. Chazono, Comput. Phys. Commun. {\bf 297}, 109058 (2024).
\bibitem{Toyokawa2013} M. Toyokawa, K. Minomo, and M. Yahiro, Phys. Rev. C {\bf 88}, 054602 (2013).
\bibitem{Negele2002} K. Amos, P. J. Dortmans, H. V. von Geramb, S. Karataglidis, and J. Raynnal, Adv. Nucl. Phys. {\bf25}, 276 (2000).
\bibitem{HFBRADcite} K. Bennaceur and J. Dobaczewski, {\it HFBRAD (v1.0)}, Comput. Phys. Commun. {\bf 168}, 96-122 (2005).
\bibitem{SKMcite} J. Bartel, P. Quentin, M. Brack, C. Guet, and H.-B. Haakansson, Nucl. Phys. A {\bf 386}, 79 (1982).
\bibitem{Franey1985} M. A. Franey and W. G. Love, Phys. Rev. C {\bf 31}, 488 (1985).
\bibitem{Per63} F. G. Perey, Direct Interactions and Nuclear Reaction Mechanism (Gordon and Breach Science Publishers, New York, 1963), p. 125.
\bibitem{Yoshida2021} K. Yoshida, Few-Body Systems {\bf62}, 28 (2021).
\bibitem{Bohr-Mottelson} A. Bohr and B. R. Mottelson, {\it Nuclear structure: single-Particle Motion and nuclear deformations}, Volume 1 (World Scientific, 1969).
\bibitem{Enciu2025}M. Enciu, A. Obertelli, P. Doornenbal, M. Heinz, W. Horiuchi, T. Inakura, W. H. Long, T. Miyagi, F. Nowacki, K. Ogata, A. Poves, A. Schwenk, K. Yoshida {\it et al.}, {\it in preparation}.
\bibitem{Tsukiyama2011} K.~Tsukiyama, S. K. Bogner, and A. Schwenk, Phys. Rev. Lett. {\bf 106}, 222502 (2011).
\bibitem{Hergert2016} H.~Hergert, S. K. Bogner, T. D. Morris, A. Schwenk and K. Tsukiyama, Phys. Rep. {\bf 621}, 165-222 (2016). 
\bibitem{Stroberg2017} S. R. Stroberg, A. Calci, H. Hergert, J. D. Holt, S. K. Bogner, R. Roth, and A. Schwenk, Phys. Rev. Lett. {\bf 118}, 032502 (2017).
\bibitem{Stroberg2019} S. R. Stroberg, H. Hergert, S. K. Bogner, and J. D. Holt, Annu. Rev. Nucl. Sci. {\bf 69}, 307-362 (2019).

\bibitem{Heinz2021} M.~Heinz, A.~Tichai, J.~Hoppe, K.~Hebeler, and A.~Schwenk, Phys.~Rev.~C {\bf 103}, 044318 (2021).
\bibitem{Stroberg2024} S. R. Stroberg, T. D. Morris, and B. C. He, Phys. Rev. C {\bf 110}, 044316 (2024).
\bibitem{Hebeler2012} K. Hebeler, S. K. Bogner, R. J. Furnstahl, A. Nogga, and A. Schwenk, Phys. Rev. C {\bf 83}, 031301(R) (2011).
\bibitem{Hagen2016}  G. Hagen, G. R. Jansen, and T. Papenbrock, Phys. Rev. Lett. {\bf 117}, 172501 (2016).
\bibitem{Taniuchi2019} R. Taniuchi, C. Santamaria, P. Doornenbal, A. Obertelli, K. Yoneda, G. Authelet, H Baba, D Calvet, F Ch\^ateau, A Corsi {\it et al.}, Nature {\bf 569}, 53–58 (2019).
\bibitem{Simonis2017} J. Simonis, S. R. Stroberg, K. Hebeler, J. D. Holt, and A. Schwenk, Phys. Rev. C {\bf 96}, 014303 (2017).
\bibitem{StrobergIMSRG} S.~R.~Stroberg, \textsc{imsrg++}, \\
https://github.com/ragnarstroberg/imsrg (2018).
\bibitem{ShimizuKSHELL} N.~Shimizu, T. Mizusaki, Y. Utsuno, and Y. Tsunoda Comput. Phys. Commun. {\bf 244}, 372-384 (2019).
\bibitem{Heinz2024} M. Heinz, T. Miyagi, S. R. Stroberg, A. Tichai, K. Hebeler, A. Schwenk, arXiv:2411.16014.

\bibitem{Aumann2021} T. Aumann, C. Barbieri, D. Bazin, C. A. Bertulani, A. Bonaccorso, W. H. Dickhoff, A. Gade, M. G\'omez-Ramos, B. P. Kay, A. M. Moro {\it et al.}, Prog. Part. Nucl. Phys. {\bf 118}, 103847 (2021).
\end{thebibliography}
\end{document}